# Design and synthesis of three-dimensional hybrid Ruddlesden-Popper nickelate single crystals

Feiyu Li,[1] Ning Guo,[2] Qiang Zheng,[2] Yang Shen,[3] Shilei Wang,[1] Qihui Cui,[1] Chao Liu,[1] Shanpeng Wang,[1] Xutang Tao,[1*] Guang-Ming Zhang,[4,5*] Junjie Zhang[1*]

[1]State Key Laboratory of Crystal Materials and Institute of Crystal Materials, Jinan, Shandong 250100, China
[2]CAS Key Laboratory of Standardization and Measurement for Nanotechnology, CAS Center for Excellence in Nanoscience, National Center for Nanoscience and Technology, Beijing, 100190 China
[3]Key Laboratory of Artificial Structures and Quantum Control, School of Physics and Astronomy, Shanghai Jiao Tong University, Shanghai 200240, China
[4]State Key Laboratory of Low-Dimensional Quantum Physics, Department of Physics, Tsinghua University, Beijing, China.
[5]Frontier Science Center for Quantum Information, Beijing 10008, China

*Email: jingti535tao@163.com (X. Tao), gmzhang@mail.tsinghua.edu.cn (G. M. Zhang), and junjie@sdu.edu.cn (J. Zhang)

**Abstract:** Advancement of technologies relies on discovery of new materials with emerging physical properties that are determined by their crystal structures. Ruddlesden-Popper (R-P) phases with formula of $A_{n+1}B_nX_{3n+1}$ ($n=1,2,3…\infty$) are among one of the most widely studied class of materials due to their electrical, optical, magnetic, thermal properties and their combined multifunctional properties.[1-6] In R-P phases, intergrowth is well-known in the short range;[7-9] however, no existing compounds have been reported to have different n mixed in bulk single crystals. Here we design a hybrid R-P nickelate $La_2NiO_4·La_3Ni_2O_7$ by alternatively stacking bilayers, which is the active structural motif in the newly discovery high-$T_c$ superconductor $La_3Ni_2O_7$ and single layers of the antiferromagnetic insulator $La_2NiO_4$. We report the successful synthesis of $La_2NiO_4·La_3Ni_2O_7$ single crystals, and X-ray diffraction and real-space imaging vis STEM show that the crystal structure consists of single layers and bilayers of $NiO_6$ octahedral stacking alternatively perpendicular to the ab plane, characterized by the orthorhombic *Immm* (No.71) space group. Resistivity measurements indicate a peculiar insulator-to-metal transition around 140 K on cooling. Correlated density functional theory (DFT+*U*) calculations corroborate this finding, and reveal that the single layer becomes paramagnetic metallic due to charge transfer via LaO layers. The discovery of $La_2NiO_4·La_3Ni_2O_7$ opens a door to access a completely new family of 3D hybrid R-P phases with the formula of $A_{n+1}B_nX_{3n+1}·A'_{m+1}B'_mX'_{3m+1}$ ($n\neq m$), which potentially host a plethora of emerging physical properties for various applications.

Quasi-2D layered R-P oxide perovskites display various emerging physical properties ranging from high temperature superconductivity,[1,10] colossal magnetoresistance,[4] metal-insulator transition[2] and multiferroicity.[5,11] The ternary oxide system La-Ni-O contains a homologous series of layered phases of general composition $(LaO)(LaNiO_3)_n$, in which $n$ layers of perovskite-type $LaNiO_3$ are separated by single NaCl-type LaO layers.[12] The average oxidation state of Ni in $La_{n+1}Ni_nO_{3n+1}$ (n=1, 2, 3, …, ∞) varies from 2+ for n=1 to 2.5+ for n=2 and to 3+ for n=∞. The first member of this series, $La_2NiO_4$, is known to have a tetragonal structure with the rotated $NiO_6$ octahedrons, being an antiferromagnetic Mott insulator at low temperatures.[13] Recently, signature of superconductivity with $T_c \sim$ 80 K above 14 GPa has reported in the second member of this series, the bilayer nickelate $La_3Ni_2O_7$.[14-17] At ambient pressure, it crystallizes in the orthorhombic *Amam* space group with Ni-O-Ni angle of 168°, and it undergoes a structural transition to *Fmmm* with Ni-O-Ni angle of 180° under high pressure.[14] So there is a nature question whether there exists other high-$T_c$ superconducting nickelates at ambient pressure, which share the similar crystal structure of $La_3Ni_2O_7$ under high pressure. Besides high-$T_c$ superconductors, are there any other types of nickelates in which rich physical properties emerge in the electronic phase diagram?

In this paper, we design a new family of perovskites by stacking $n$ and $m$ layer perovskites in long-range order, yielding three-dimensional $A_{n+1}B_nO_{3n+1} \cdot A'_{m+1}B'_mO_{3m+1}$ hybrid oxide perovskites. As an example, we successfully synthesize single crystals of $La_2NiO_4 \cdot La_3Ni_2O_7$ (n=1, m=2) for the first time. The key structural motif for the electronic properties in $La_3Ni_2O_7$ is the bilayer Ni-O planes. On average, each $Ni^{2.5+}$ is half-filled in its $3d_{z^2}$ orbital and quarter-filled in its $3d_{x^2-y^2}$ orbital.[14] Incorporation of the bilayer $NiO_2$ into a new compound is expected to maintain the key electronic features of $La_3Ni_2O_7$. In $La_2NiO_4$, the single-layer $NiO_2$ planes form a two-dimensional spin-1 Heisenberg antiferromagnetic insulator on a square lattice with a Neel temperature of ~330 K.[13] Hole doping in $La_2NiO_4$ results in rich electronic phase diagram, consisting of antiferromagnetic insulator, charge and spin stripes, checkboard charge order and pseudogap.[18] **Fig. 1a** presents our materials design strategy to hybridize bilayer from $La_3Ni_2O_7$ and single layer from $La_2NiO_4$ in the long-range order perpendicular to the Ni-O plane. By tailoring $La_2NiO_4$ into single layers, the antiferromagnetic long-range order is expected to be destroyed and only spin fluctuations are left. The combination of magnetic single layer and superconducting bilayers are predicted to host emerging properties that belong to neither $La_2NiO_4$ nor $La_3Ni_2O_7$.

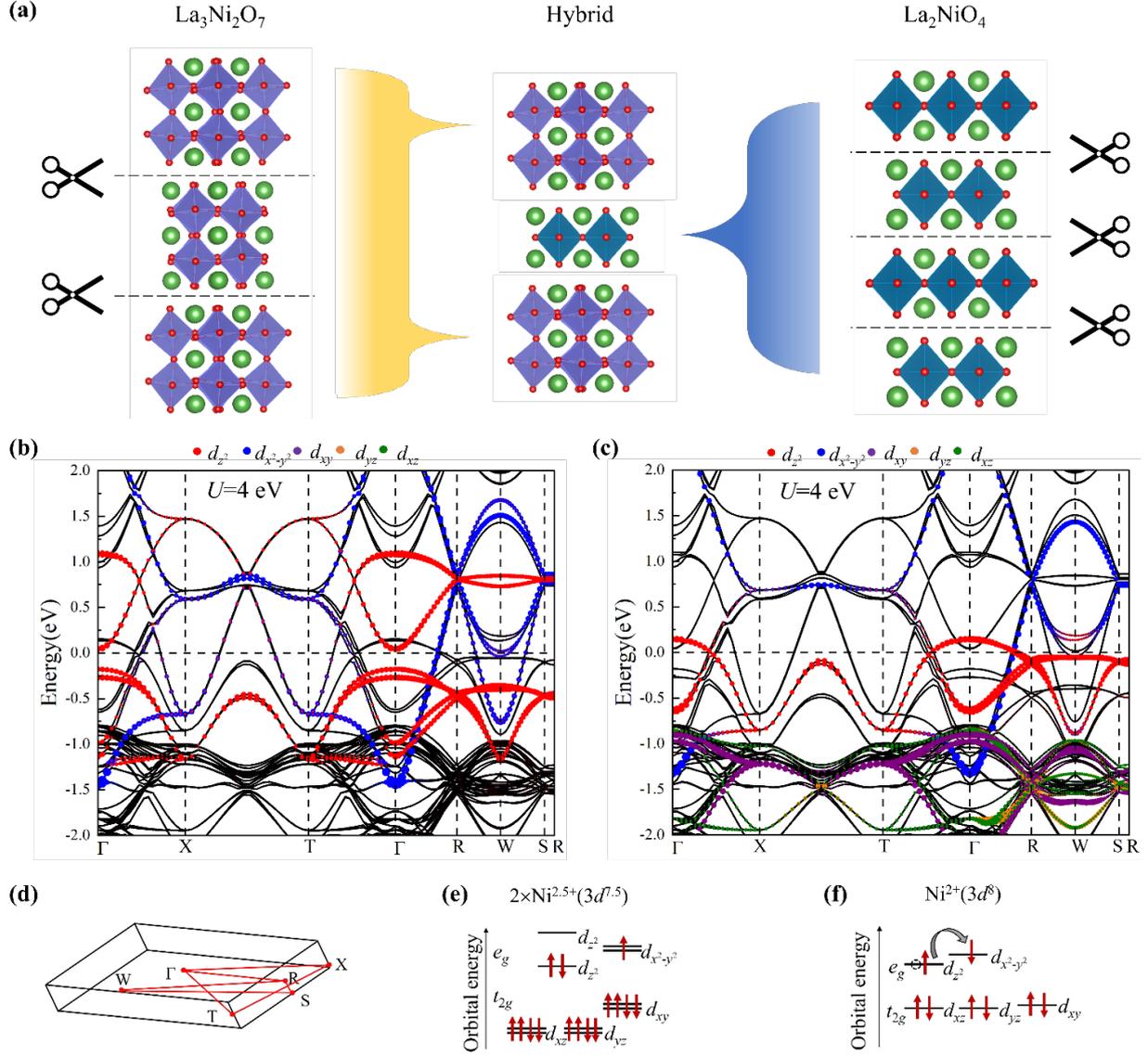

**Figure 1. Materials design strategy for hybrid La₂NiO₄·La₃Ni₂O₇ and calculated electronic structure.** (a) Illustration of our design strategy to hybridize bilayer, the structural motif that is responsible for high-temperature superconductivity in La$_3$Ni$_2$O$_7$, and single layer in the 3D antiferromagnetic La$_2$NiO$_4$ in the long-range order. (b) Orbital projected weights of the bilayer NiO$_2$ in the hybrid La$_2$NiO$_4$·La$_3$Ni$_2$O$_7$ crystal given by the correlated density functional theory. (c) Orbital projected weights of the single NiO$_2$ in the hybrid La$_2$NiO$_4$·La$_3$Ni$_2$O$_7$ crystal given by the correlated density functional theory. (d) Schematic of the 3D body-centered orthorhombic Brillouin zone. The red lines correspond to the paths of the electronic bands in (b) and (c). (e) Energy levels of the unit with two Ni$^{2.5+}$ (3d$^{7.5}$) in the bilayer under the crystal field. The $d_{z^2}$ orbitals of two Ni cations in the neighboring layers form the bonding and anti-bonding states, (f) Energy levels of the unit with Ni$^{2+}$ (3d$^8$) in the single layer under the crystal field. The dotted circle indicates a hole generated by charge transfer from $d_{z^2}$ to $d_{x^2-y^2}$ orbitals.

With this design, we use flux method to synthesize the single crystals of $La_2NiO_4 \cdot La_3Ni_2O_7$ at ambient pressure. Conventionally, single crystals of rare earth nickelates with higher nickel valence are prepared either by high oxygen pressure floating zone techniques or high pressure flux method.[12,19] Recent advancement in growing single crystals of trilayer $La_4Ni_3O_{10}$ has been achieved without high pressure, i.e., single crystals can be prepared at ambient pressure using $K_2CO_3$ as a flux.[20] By changing the initial molar ratio of La and Ni to 5:3, single crystals of $La_2NiO_4 \cdot La_3Ni_2O_7$ with dimensions up to 120 μm in length and 80 μm in thickness were prepared. **Fig. 2a** presents the SEM image of a typical single crystal with sizes of 110×80×50 μm$^3$ (more crystals are shown in **Fig. S1**). The regular shape with smooth and clean surface of the sample indicates high quality of the as-grown single crystal. EDS measurements on single crystals result in La:Ni of 1.77 (sample 1) and 1.70 (sample 2), consistent with the expected 5:3 for $La_2NiO_4 \cdot La_3Ni_2O_7$.

The crystal structure at room temperature was determined using in-house X-ray single crystal diffraction. Initial unit cell determination resulted in a unit cell with $c$~16.5 Å, which lies in the middle of $La_2NiO_4$ with $c$~12.7 Å[21] and $La_3Ni_2O_7$ with $c$~20.5 Å.[14] This observation strongly suggests that it is a long-range intergrowth of single layer and bilayer $NiO_2$ planes. Subsequent analysis (**Fig. S2**) leads to orthorhombic body centered unit cell with lattice parameters of a= 5.4296(2) Å, b=5.4330(2) Å and c= 33.2067(12) Å. **Figs. 2b & 2c** display selected high-symmetry reciprocal lattice planes ($hk0$) and ($h0l$) reconstructed from a total of 2728 frames. The observed peaks obey the selection rule of $h+k+l$=even, consistent with the body centered unit cell. The structure was solved using the *Immm* (No.71) space group, and the refinements converged to $R_1$=6.05% and $wR_2$=10.23%. **Fig. S3** shows the observed structural factor F as a function of the calculated values. Ideally the calculated F equals to the observed F as indicated by the red line. Details of crystal parameters, data collection, and structure refinement are summarized in **Table S1**.

**Fig. 2d** presents the crystal structure of $La_2NiO_4 \cdot La_3Ni_2O_7$ using the ball-and-stick model. The unit cell contains two bilayers and two single-layers. The asymmetric unit consists of six La atoms, three Ni atoms and eight O atoms. The La atoms are surrounded by nine or thirteen oxygen atoms with bond length in the range of 2.28(5) - 2.778(11) Å. The Ni atoms are bonded to six oxygen atoms with bond length of 1.916(10) - 2.227(18) Å for the bilayer (**Fig. 2e**) and 1.92026(5) - 2.34(5) Å for the single layer (**Fig. 2f**). The Ni-O bond distance along $c$ axis for the bilayer is 3% - 6.5% longer than that of $La_3Ni_2O_7$ or $La_2PrNi_2O_7$ measured at 14-15 GPa.[14,22] The out-of-plane Ni-O-Ni angle within the bilayer is 180° (**Fig. 2e**), as the same as that of $La_3Ni_2O_7$ in the superconducting state under high pressure.[14] These crystallographic data suggest that $La_2NiO_4 \cdot La_3Ni_2O_7$ is a potential material for exploring high temperature superconductivity.

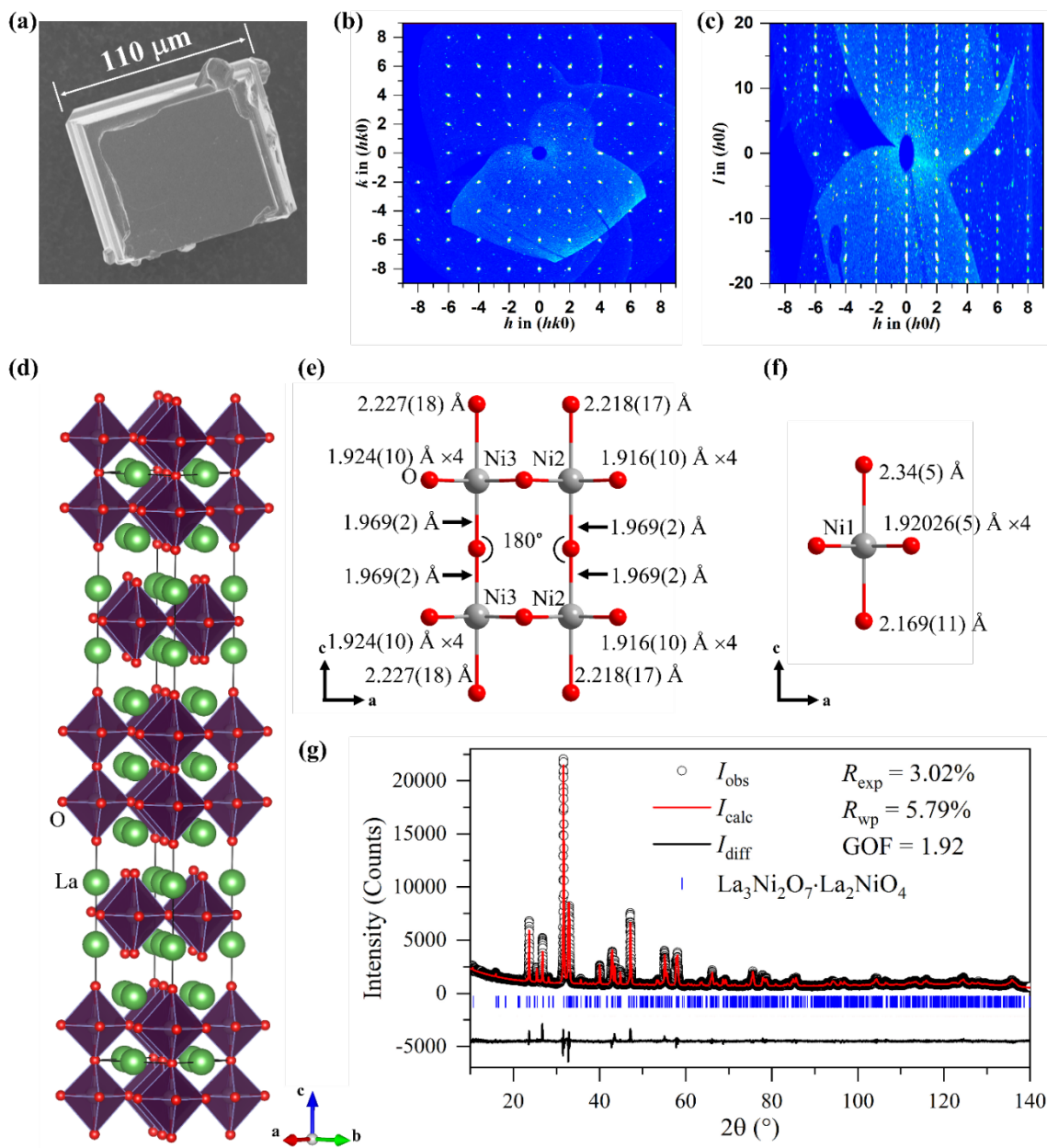

**Figure 2. Crystal structure of the single-layer and bilayer hybridized nickelate La$_2$NiO$_4$·La$_3$Ni$_2$O$_7$.** **(a)** A SEM image of an as-grown single crystal with dimensions of 110×80×50 μm$^3$. **(b)** Reconstructed (*hk*0) planes from in-house X-ray single crystal diffraction data collected at 296(2) K. **(c)** Reconstructed (*h*0*l*) planes from in-house X-ray single crystal diffraction data collected at 296(2) K. **(d)** Structural model obtained from X-ray single crystal diffraction. **(e)** Ball-and-stick drawings of the NiO$_6$ octahedrons with bond distances and bond angles in the bilayer NiO$_2$. **(f)** Ball-and-stick drawing of the NiO$_6$ octahedrons with bond distances in the single layer NiO$_2$. **(g)** Rietveld refinement on powder diffraction data collected on pulverized single crystals obtained from flux growth. Structural model in **(d)** was used as a starting model.

Rietveld refinement on powder diffraction data using the single crystal model was carried out to verify our structural model (**Fig. 2g**). The refinement converged to the following figures of merit: $R_{exp}$=3.02%, $R_{wp}$=5.79%, and GOF=1.92 with lattice parameters of $a$= 5.41730(9) Å, $b$= 5.45820(9) Å, $c$= 33.1844(6) Å. These reasonable parameters and the small difference between calculated and observed intensity strongly support the crystal structure from single crystal diffraction. We found that there exists a small amount of $La_3Ni_2O_7$ (6.7 wt %) impurity (not shown in the figure) with the rest 93.3 wt % attributable to $La_2NiO_4 \cdot La_3Ni_2O_7$. The powders were prepared by pulverizing crystals that previously did not pass through 300-mesh sieves (crystals >50 μm on edge). There exist two possibilities for the source of the impurity: (1) single crystals of $La_3Ni_2O_7$, and (2) inclusions or intergrowth of $La_3Ni_2O_7$ on $La_2NiO_4 \cdot La_3Ni_2O_7$. We performed single crystal X-ray diffraction experiments multiple times to select $La_2NiO_4 \cdot La_3Ni_2O_7$ single crystals, and we occasionally picked up $La_3Ni_2O_7$ single crystals. By further considering the clean single crystal diffraction pattern and STEM results (show below), the first source is likely the case. It is difficult to distinguish $La_2NiO_4 \cdot La_3Ni_2O_7$ from $La_3Ni_2O_7$ single crystals using microscope due to their similar morphology.

To further verify the long-range intergrowth of single layer and bilayer $NiO_2$ planes, real-space imaging via STEM was performed. A typical HAADF-STEM image in the [110] projection shown in **Fig. 3a** and a lower-magnification HAADF image shown in **Fig. S4** indicate perfectly ordered stacking of the alternating bilayers and single layers in the length scale of tens of nanometers in the $La_2NiO_4 \cdot La_3Ni_2O_7$ single crystals. Such a long-range ordered stacking can be revealed by chemical distributions of La and Ni from EDS maps in **Fig. 3b**. As shown in **Fig. S5** and **Fig. S6**, the HAADF-STEM imaging was also performed along the [100] zone axis, further confirming the fully ordering sequences of the alternating bilayers and single layers.

**Fig. 3c** shows the magnetic susceptibility of the as-grown $La_2NiO_4 \cdot La_3Ni_2O_7$ single crystals in the temperature range of 25-350 K under an external magnetic field of 0.4 T. Between 25 and ~200 K, it exhibits the Curie-Weiss behavior, indicating the presence of weakly interacting local magnetic moments. Above ~200 K, the magnetic susceptibility is almost temperature independent, consistent with Pauli paramagnetism. The data (not shown) between 2 and 25 K does not show any magnetic long-range order but strong sample dependent. **Fig. 3d** displays the resistivity of polycrystalline pellet of $La_2NiO_4 \cdot La_3Ni_2O_7$ pulverized from single crystals and sintered at 900 °C in air. In addition to the expected metal-insulator transition at high temperature above 350 K, a peculiar insulator to metal transition occurs around 140 K on cooling, in sharp contrast to the other R-P nickelates such as the insulating behavior of $La_2NiO_4$ at low temperatures,[13] the metal-to-metal transition in bilayer $La_3Ni_2O_7$ and trilayer $La_4Ni_3O_{10}$,[23-25] and the metal-to-insulator transition in $RNiO_3$ (R=Pr - Lu, Y).[26] The underlying physics of this insulator-to-metal transition may be related to the presence of charge transfer between insulating single layer and metallic bilayers in this hybrid structure.

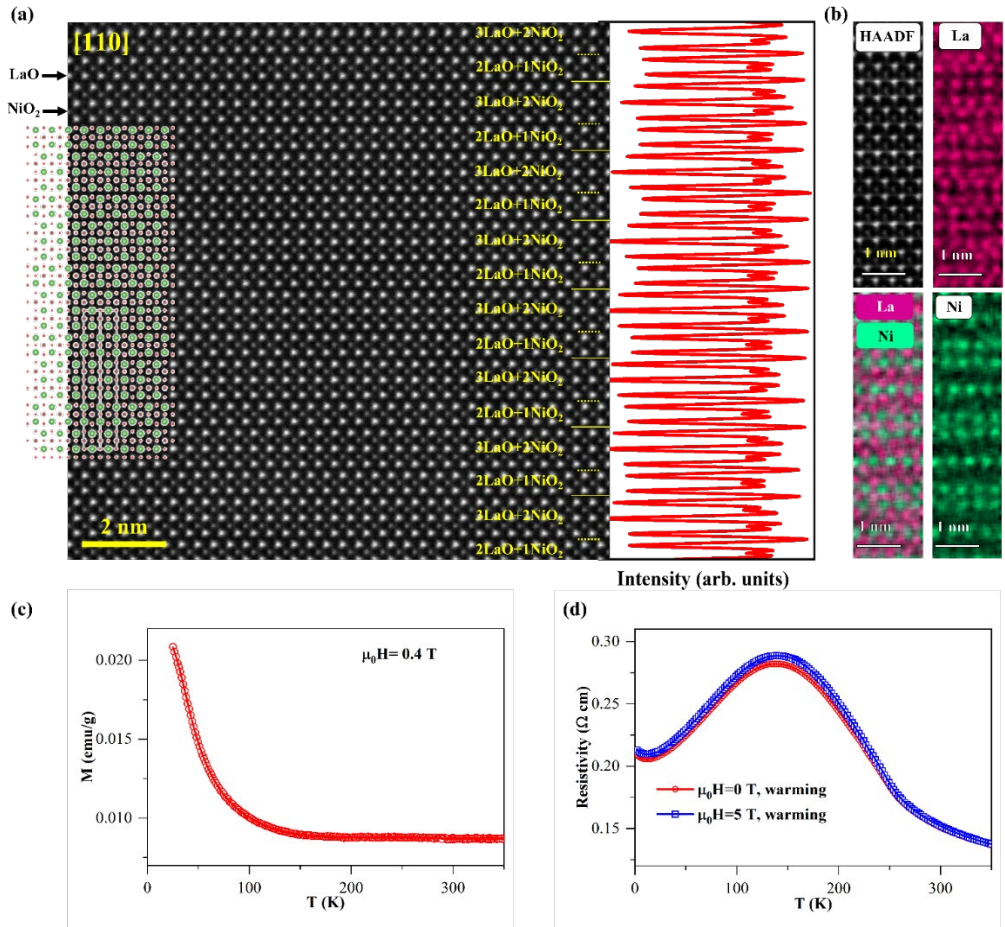

**Figure 3. Real-space imaging of the structure and physical properties of $La_3Ni_2O_7 \cdot La_2NiO_4$. (a)** A typical atomic-scale HAADF-STEM image in the projection of [110] with overlaid crystal structure model; the right panel is line intensity profile for rows of all atomic columns. **(b)** EDS maps for La and Ni and mixed color map of them. **(c)** Magnetic susceptibility of as-grown single crystals. **(d)** Resistivity on warming under an external magnetic field of 0 and 5 T. Note a single crystal is too small to put four leads on, polycrystalline pellets were used and treated at 900 °C in air prior to making contacts.

With the determined structure of single crystals and low-temperature physical properties, we performed the numerical calculation of correlated density functional theory (DFT+$U$) for the single crystal $La_2NiO_4 \cdot La_3Ni_2O_7$ within a non-magnetic solution. With an intermediate on-site Coulomb interaction $U$=4 eV, we find that the electronic states near the Fermi level are dominated by the Ni $3d_{x^2-y^2}$ and $3d_{z^2}$ orbitals in the energy window of -2 eV to 2 eV, while the Ni $t_{2g}$ orbital electrons are far below the Fermi energy (**Fig. S7**). The Ni $3d_{z^2}$ orbitals are mixed with oxygen $2p_z$ orbitals across the

Fermi level, indicating strong covalent hybridization among these orbitals (**Fig. S7**). The rare-earth element La has little hybridization with Ni and O $2p_z$ orbitals (**Fig. S8**). Actually, it had better to present complementary plots of the Ni $e_g$ bands disentangled according to their origin from Ni sites in the different structural units. **Fig. 1b** displays the orbital projected weights of the structural units with $NiO_6$ octahedral bilayers in the hybrid $La_2NiO_4 \cdot La_3Ni_2O_7$ crystal, where the $3d_{z^2}$ orbitals form the σ-bonding and anti-bonding bands via the inter-layer apical oxygens, opening a small energy gap at the Fermi energy (**Fig. 1d**). This gives rise to the half-filled $3d_{z^2}$ orbitals and the quarter filled $3d_{x^2-y^2}$ orbitals, very similar to the electronic structure of $La_3Ni_2O_7$.[14] However, the orbital projected weights of the structural units with $NiO_6$ octahedral single layers in the hybrid $La_2NiO_4 \cdot La_3Ni_2O_7$ crystal show that the $3d_{x^2-y^2}$ electrons are nearly free with parabolic dispersions along the Γ-R direction, the $3d_{z^2}$ orbital band is hole doped with a narrow bandwidth, and the van Hove singularity of correlated $3d_{z^2}$ electrons along the R-S direction may induce a possible electronic instability (**Fig. 1c**). These numerical results can be simplified as shown in **Figs. 1e & 1f**, and are consistent with the paramagnetic metallic behavior observed in the low-temperature experimental result.

Therefore, our proposed long-range hybrid single layer and bilayer $NiO_2$ planes in a single compound have been realized in $La_2NiO_4 \cdot La_3Ni_2O_7$, as evidenced by X-ray single crystal diffraction, Rietveld refinement on powder diffraction, and real-space imaging of the structure using STEM. In fact, this new compound just represents one member of a new family of long-range hybrid R-P nickelates. **Fig. 4a** shows the conventional R-P nickelates with the formula of $La_{n+1}Ni_nO_{3n+1}$ (n=1, 2, 3, …∞).[27] By expanding this R-P nickelate via introducing different building blocks, the hybrid R-P family of nickelates $La_{n+1}Ni_nO_{3n+1} \cdot La_{m+1}Ni_mO_{3m+1}$ (m, n=1, 2, 3… and m>n) can be formed as shown in **Fig. 4b**. Among them, $La_2NiO_4 \cdot La_3Ni_2O_7$ is the first member of this family with n=1 and m=2. By tuning conditions of materials synthesis during flux growth[20] or high pressure floating zone growth,[24] it is expected that $La_2NiO_4 \cdot La_4Ni_3O_{10}$ (n=1, m=3), $La_2NiO_4 \cdot La_5Ni_4O_{13}$ (n=1, m=4), $La_2NiO_4 \cdot La_6Ni_5O_{16}$ (n=1, m=5), $La_3Ni_2O_7 \cdot La_4Ni_3O_{10}$ (n=2, m=3), $La_3Ni_2O_7 \cdot La_5Ni_4O_{13}$ (n=2, m=4), $La_3Ni_2O_7 \cdot La_6Ni_5O_{16}$ (n=2, m=5), $La_4Ni_3O_{10} \cdot La_5Ni_4O_{13}$ (n=3, m=4), $La_4Ni_3O_{10} \cdot La_6Ni_5O_{16}$ (n=3, m=5), $La_5Ni_4O_{13} \cdot La_6Ni_5O_{16}$ (n=4, m=5), … can be synthesized.

Even more nickelates can be designed and explored if one mixes different rare earth ions. Synthesis of these bulk samples may be challenging for $n$, $m$>3. As an alternative way, the preparation of thin films using techniques like molecular-beam epitaxy[28,29] potentially have a better chance. **Fig. 4c** shows the general R-P phases $A_{n+1}B_nX_{3n+1} \cdot A'_{m+1}B'_mX'_{3m+1}$, where A and A' are usually alkaline earth or rare earth metal ion (e.g. Ca, Sr, Ba, La-Lu) or organic groups (e.g., $CH_3NH_3$, $CH(NH_2)_2$), B and B' are transition metal ion (e.g. Ti, V, Mn, Fe, Co, Ni, Nb, Zr, Ru, Sn, Pb), X and X' are anion (e.g. O, Cl,

Br, I) and *m, n*=1, 2, 3,…∞. With the success of $La_2NiO_4 \cdot La_3Ni_2O_7$ in the nickelate system, it is straightforward to explore more general hybrid R-P phases in other transition metal oxides including cobaltites for spin transition,[30] manganites for colossal magnetoresistance,[4] halide perovskites for photovoltaic aplications[3] and X-ray or γ-ray detection.[31,32] Indeed, a superlattice of $Sr_2IrO_4 \cdot Sr_3Ir_2O_7$ with dimensions of a few microns in thickness, thicker than layer-by-layer grown thin films but thinner than typical bulk crystals, have been achieved,[33] and synthesis of bulk single crystals of this material and characterization of its physical properties are intriguing.

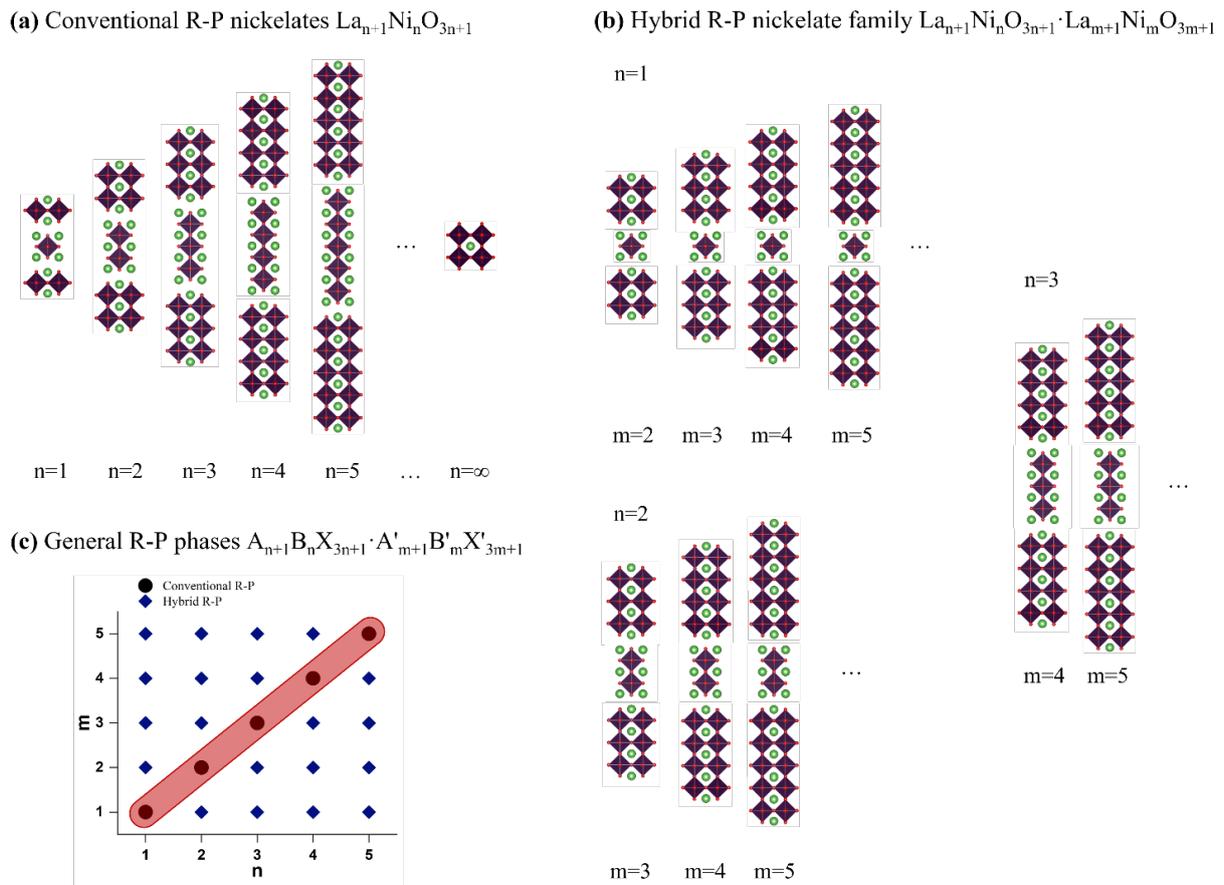

**Figure 4. Hybrid Ruddlesden-Popper (R-P) phases as a new class of quantum materials that potentially host emerging physical properties. (a)** Conventional R-P nickelates with the formula of $La_{n+1}Ni_nO_{3n+1}$ (n=1, 2, 3…). Note n are integers and only one n exists in a single compoud. **(b)** Hybrid R-P nickelate family with the formula of $La_{n+1}Ni_nO_{3n+1} \cdot La_{m+1}Ni_mO_{3m+1}$ (m, n=1, 2, 3… and m≠n). $La_2NiO_4 \cdot La_3Ni_2O_7$ is one member of this family with n=1 and m=2. **(c)** General R-P phases with the formula of $A_{n+1}B_nX_{3n+1} \cdot A'_{m+1}B'_mX'_{3m+1}$ (m, n=1, 2, 3…). In these compounds, A and A' are usually alkaline earth or rare earth metal ion (e.g. Ca, Sr, Ba, La-Lu) or organic groups (e.g., $CH_3NH_3$, $CH(NH_2)_2$); B and B' are transition metal ion (e.g. Ti, V, Mn, Fe, Co, Ni, Nb, Zr, Ru, Sn, Pb); X and X' are anion (e.g. O, Cl, Br, I).

Furthermore, in the trilayer or bilayer R-P nickelates, the apical oxygen atoms between $NiO_2$ layers can be removed upon topotactical reduction,[34,35] but the in-plane oxygen atoms remain unchanged, making $La_2NiO_4 \cdot La_3Ni_2O_6$ or $La_2NiO_4 \cdot La_4Ni_3O_8$ a unique system to explore R-P and the Ni-O planar (T'-structure) hybrid phase. For other members of the hybrid R-P nickelates (n>1, m>n), if succeed, it will be interesting to perform topotactical reduction to hybrid T'-structure nickelates for rich emergent physical properties such as charge order and high-$T_c$ superconductivity.[36]

In summary, by tailoring the bilayer $NiO_2$ from the recently discovered high-$T_c$ superconductor $La_3Ni_2O_7$ and the single layer from the antiferromagnetic insulator $La_2NiO_4$, we have designed a long-range hybrid R-P nickelate $La_2NiO_4 \cdot La_3Ni_2O_7$. Realization of such a structure in real materials has been achieved via flux growth at ambient pressure. The crystal structure of $La_2NiO_4 \cdot La_3Ni_2O_7$ consists of single layers and bilayers stacking alternatively perpendicular to the ab plane as determined from single crystal X-ray diffraction. The structure is corroborated by Rietveld refinement on powder diffraction and confirmed by real-space imaging using STEM. This new material shows the Curie-Weiss behavior as indicated by magnetic susceptibility and a novel insulator-to-metal transition at low temperatures that has never been reported in nickelates. The discovery of $La_2NiO_4 \cdot La_3Ni_2O_7$ opens up horizons to a completely new family of hybrid R-P phases with the formula of $A_{n+1}B_nX_{3n+1} \cdot A'_{m+1}B'_mX'_{3m+1}$ (n ≠ m), which potentially host a plethora of emerging physical properties including high temperature superconductivity, colossal magnetoresistance, metal-insulator transition, multiferroicity and optoelectronic properties.

**Note**. During the preparation of this manuscript, we are informed that Chen et al. from Argonne National Laboratory successfully synthesized $La_2NiO_4 \cdot La_4Ni_3O_{10}$ (n=1, m=3) crystals and reported its crystal structure and resistivity measurement.[37] The existence of this nickelate phase is also reported by Puphal et al. by Germany group independently.[38] This so-called "1313" compound just serves as the second example of our proposed hybrid R-P family of nickelates.

**Acknowledgements**

Work at Shandong University was supported by the National Natural Science Foundation of China (12074219 and 12374457), the 111 Project 2.0 (BP2018013), the TaiShan Scholars Project of Shandong Province (tsqn201909031), the QiLu Young Scholars Program of Shandong University, the Crystalline Materials and Industrialization Joint Innovation Laboratory of Shandong University and Shandong Institutes of Industrial Technology (Z1250020003), and the Project for Scientific Research Innovation Team of Young Scholars in Colleges and Universities of Shandong Province (2021KJ093). G. M. Zhang acknowledges the support of China's Ministry of Science and Technology (Grant No. 2023YFA1406400). J. Z. and F. L. thank Prof. Jian Zhang for his help with in-house single crystal X-ray diffraction. J.Z. thanks Dr. Yu-Sheng Chen and Dr. Tieyan Chang from The University of Chicago for stimulating discussions.

**Author contributions**

J.Z. and X.T. conceived the project and G.M. Z proposed the physical picture to design the hybrid RP family of materials. F.L. grew single crystals with the assistance of C.L. F.L. performed the powder and single-crystal X-ray diffraction experiments. F.L. performed SEM, EDS, magnetic susceptibility, and transport measurements with the help of Q.C. S.L.W., S.P.W. and C.L. N.G. and Q.Z. carried out STEM measurements and performed data analysis. F.L., J.Z., and G.M.Z. analyzed data. Y.S. performed DFT calculations with helps from G.M. Z. J.Z. and G.M. Z. wrote the manuscript with contributions from all coauthors.

**Competing interests**

The authors declare no conflict of interest.

**Experimental Section**

**Single crystal growth.** Single crystals of La$_2$NiO$_4$·La$_3$Ni$_2$O$_7$ were grown for the first-time using flux method with K$_2$CO$_3$ as a flux at ambient pressure. Mixture of preheated La$_2$O$_3$ (Sigma-Aldrich, 99.99%), Ni (Alfa Aesar, 99.8%, particle size 5-15 μm) and anhydrous K$_2$CO$_3$ (Aladdin, 99.99%) in the molar ratio of 5:6:210 were loaded in a Al$_2$O$_3$ crucible with a lid in order to minimize the evaporation. The procedure for crystal growth is similar to that of La$_4$Ni$_3$O$_{10}$.[20]

**X-ray diffraction.** Single crystal X-ray diffraction data were collected using a Bruker AXS D8 Venture (Mo-$K\alpha_1$ radiation, λ = 0.71073 Å) diffractometer at room temperature. A single crystal with dimensions of 40×57×58 μm$^3$ was used to determine the structure of La$_2$NiO$_4$·La$_3$Ni$_2$O$_7$. For data collection, Ω and Φ scans were used, and 2728 frames were collected. Indexing was performed using Bruker APEX4 software.[39] Data integration and cell refinement were performed using SAINT, and multiscan absorption corrections were applied using the SADABS program.[39] Using Olex2,[40] the structure was solved with the XT structure solution program using Intrinsic Phasing and refined with the XL refinement package using Least Squares minimisation.[39] All La and Ni atoms (except O atoms) were modeled using anisotropic ADPs, and the refinements converged for $I > 2\sigma$ ($I$), where $I$ is the intensity of reflections and $\sigma(I)$ is standard deviation. Further details of the crystal structure investigations may be obtained from the joint CCDC/FIZ Karlsruhe online deposition service by quoting the deposition number CSD 2313044. Powder X-ray powder diffraction data were collected on a Bruker AXS D2 Phaser diffractometer at room temperature using Cu-$K\alpha$ radiation (λ = 1.5418 Å) in the 2θ range of 10-140° with a scan step size of 0.01° and a scan time of 2.1 s per step. TOPAS 6 was used for Rietveld refinement where single-crystal structural model was used as a starting model. Refinement parameters include background (chebychev function, order 5), sample displacement, lattice parameters, crystallite size L and strain G.

**Scanning Electron Microscopy (SEM).** The morphology of the as-grown crystals was examined using a scanning electron microscope. The scanning electron microscope images were obtained by Hitachi S-4800 microscope incident electron of 5.0 kV.

**Energy Dispersive Spectrometer (EDS).** The X-ray spectrometer EDAX GENESIS XM2 SYSTEM 60x on S-4800 was used for qualitative and quantitative analysis of the as-grown crystals. The experiments were carried out on three single crystals.

**Scanning transmission electron microscopy (STEM).** STEM specimens were prepared by crushing La$_2$NiO$_4$·La$_3$Ni$_2$O$_7$ single crystals in ethanol. A dop of the suspensions was deposited on lacey carbon-coated copper grids and dried in air. High-angle annular dark-field (HAADF)-STEM images were acquired at an accelerating voltage of 300 kV on a double-aberration-corrected transmission electron microscope (Spectra 300, Thermo Fisher Scientific), equipped with a field-emission electron source. The probe convergence semi-angle and inner collection seimi-angle are 25.0 mrad and 49 mrad, respectively. Energy dispersive spectroscopy (EDS) data were obtained with a Super-X EDS detector.

**Resistivity.** The resistivity of $La_2NiO_4 \cdot La_3Ni_2O_7$ was measured using the standard four-probe method on polycrystalline pellets under an external magnetic field of 0 and 5 T. Note a single crystal is too small to put four leads on, polycrystalline pellets were used and treated at 900 °C in air prior to making contacts with silver paste. A Quantum Design Physical Property Measurement System (DynaCool-9) was used to measure resistivity in the temperature range of 2-300 K at a warming rate of 3.0 K/min.

**Magnetic susceptibility.** DC magnetization of as-grown $La_2NiO_4 \cdot La_3Ni_2O_7$ single crystals was measured using a PPMS DynaCool 9 T. ZFC-W (zero-field cooling with data collected on warming), FC-C (field cooling and data collected on cooling), and FC-W (field cooling and data collected on warming) data were collected between 2 and 300 K under an external magnetic field of 0.4 T. After cooled to 2 K under zero magnetic field, data were collected on warming at a rate of 3 K/min. FC-C and FC-W data collection used the same rate of 3 K/min.

**DFT calculations.** The first-principles calculations were performed with the density functional theory using the projector augmented wave (PAW) method[41] implemented in the Vienna Ab Initio Simulation Package (VASP).[42,43] The exchange correlation potential is described by generalized gradient approximation (GGA) of Perdew-Burke-Ernzerhof (PBE) functions.[44] Strong electron-electron correlation beyond GGA for nickel $3d$ electrons is supplemented by plus Hubbard $U$ (GGA + $U$) calculations with Dudarev's approach,[45] and we take $U = 4$ eV in this work. An energy cutoff of 600 eV for the plane-wave expansion and a 7×6×1 Monkhorst-Pack grid for $k$-point sampling are adopted for self-consistent calculations with good convergence (with accuracy at $10^{-6}$ eV). The crystal structure was fixed to the experimentally refined lattice constants obtained from SXRD.

Supplementary Materials for:

# Design and synthesis of three-dimensional hybrid Ruddlesden-Popper nickelate single crystals

Feiyu Li,[1] Ning Guo,[2] Qiang Zheng,[2] Yang Shen,[3] Shilei Wang,[1] Qihui Cui,[1] Chao Liu,[1] Shanpeng Wang,[1] Xutang Tao,[1*] Guang-Ming Zhang,[4,5*] Junjie Zhang[1*]

[1]State Key Laboratory of Crystal Materials and Institute of Crystal Materials, Jinan, Shandong 250100, China

[2]CAS Key Laboratory of Standardization and Measurement for Nanotechnology, CAS Center for Excellence in Nanoscience, National Center for Nanoscience and Technology, Beijing, 100190 China

[3]Key Laboratory of Artificial Structures and Quantum Control, School of Physics and Astronomy, Shanghai Jiao Tong University, Shanghai 200240, China

[4]State Key Laboratory of Low-Dimensional Quantum Physics, Department of Physics, Tsinghua University, Beijing, China.

[5]Frontier Science Center for Quantum Information, Beijing 10008, China

*Email: jingti535tao@163.com (X. Tao), gmzhang@mail.tsinghua.edu.cn (G. M. Zhang), and junjie@sdu.edu.cn (J. Zhang)

**Figure S1.** A photo of as-grown single crystals of $La_2NiO_4 \cdot La_3Ni_2O_7$ using flux method.

**Figure S2.** Initial determination of unit cell using APEX3.

**Figure S3.** Calculated structural factor using the Immm model as a function of observed structural factor from the single crystal X-ray diffraction data.

**Table S1.** Crystal data and structure refinement for $La_2NiO_4 \cdot La_3Ni_2O_7$.

**Figure S4.** Real-space imaging of $La_2NiO_4 \cdot La_3Ni_2O_7$ in the projection of [110].

**Figure S5.** Real-space imaging of $La_2NiO_4 \cdot La_3Ni_2O_7$ in the projection of [100].

**Figure S6.** Real-space imaging of $La_2NiO_4 \cdot La_3Ni_2O_7$ in the projection of [100] with lower-magnification.

**Figure S7**. DFT+$U$ band structure calculations of non-magnetic $La_2NiO_4 \cdot La_3Ni_2O_7$. Orbital projected weights from all Ni 3d orbitals and all O 2p orbitals.

**Figure S8**. DFT+$U$ band structure calculations of non-magnetic $La_2NiO_4 \cdot La_3Ni_2O_7$. Orbital projected weights from La 5d orbitals compared to O 2p orbitals, indicating the presence of charge transfer via the LaO chains between the bilayer and monolayer of nickel oxide planes within the unit cell.

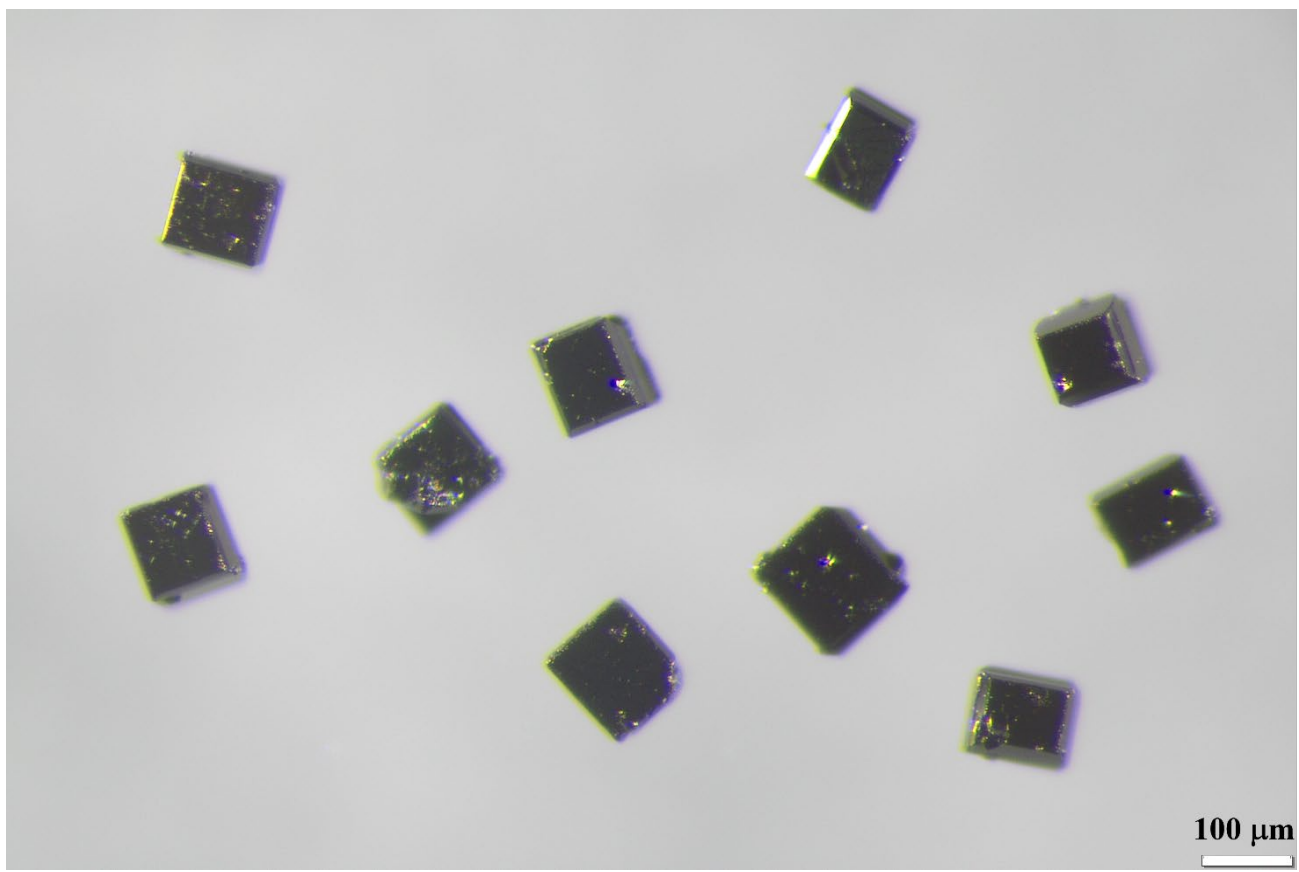

**Figure S1. A photo of as-grown single crystals of La$_2$NiO$_4$·La$_3$Ni$_2$O$_7$ using flux method.** The typically dimensions of single crystals are 100-120 μm in length and width and 30-50 μm in thickness.

**(a)** a*b* plane in the reciprocal space
$a$=3.84 Å, $b$=3.84 Å, $c$=16.55 Å, $\alpha$=89.98°, $\beta$=89.99°, $\gamma$=90.06°

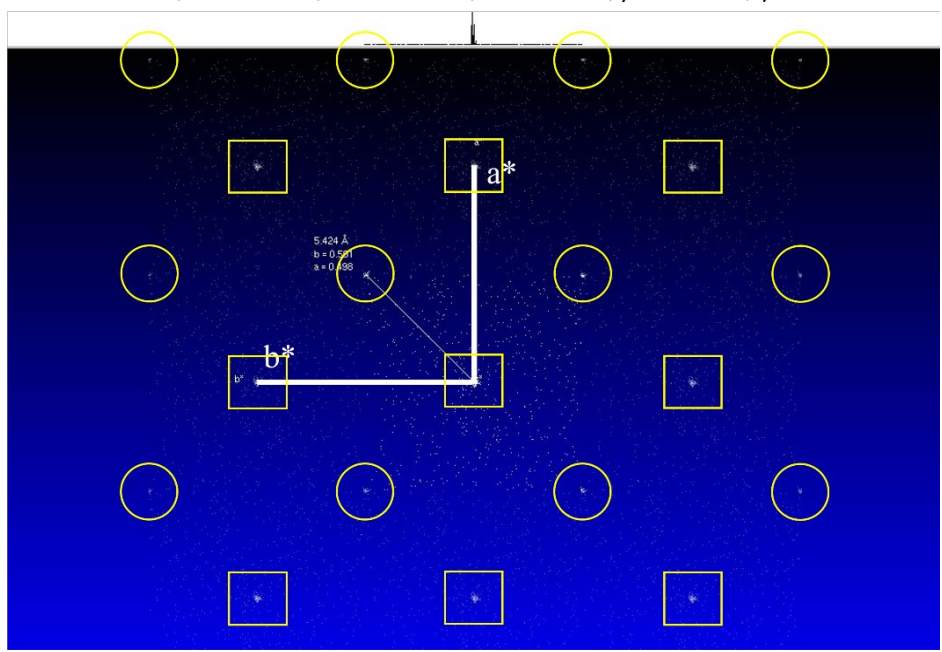

**(b)** a*c* plane in the reciprocal space

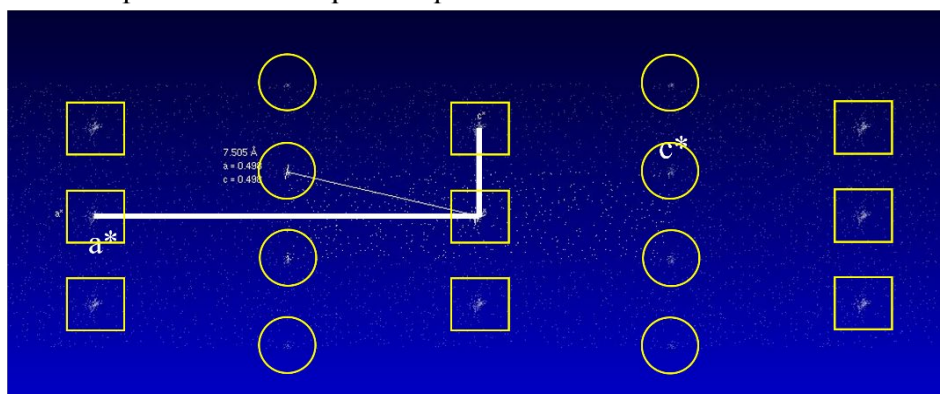

**Figure S2. Initial determination of unit cell in the reciprocal space using APEX3. (a)** Reflections with a wavevector of (1/2, 1/2, 0) as marked by circles are clearly seen using the orthorhombic setting with $a$~3.83 Å, $b$~3.83 Å and $c$~16.55 Å. This indicates the unit cell should be twice large, i.e., a~5.4 Å, b~5.4 Å. **(b)** Reflections with a wavevector of (1/2, 0, 1/2) are clearly seen using the orthorhombic setting with a~b~3.83 and c~16.55. This indicates the $c$ axis should be doubled. Finally, the correct unit cell is orthorhombic with $a$= 5.4296(2) Å, $b$= 5.4330(2) Å, $c$= 33.2067(12) Å.

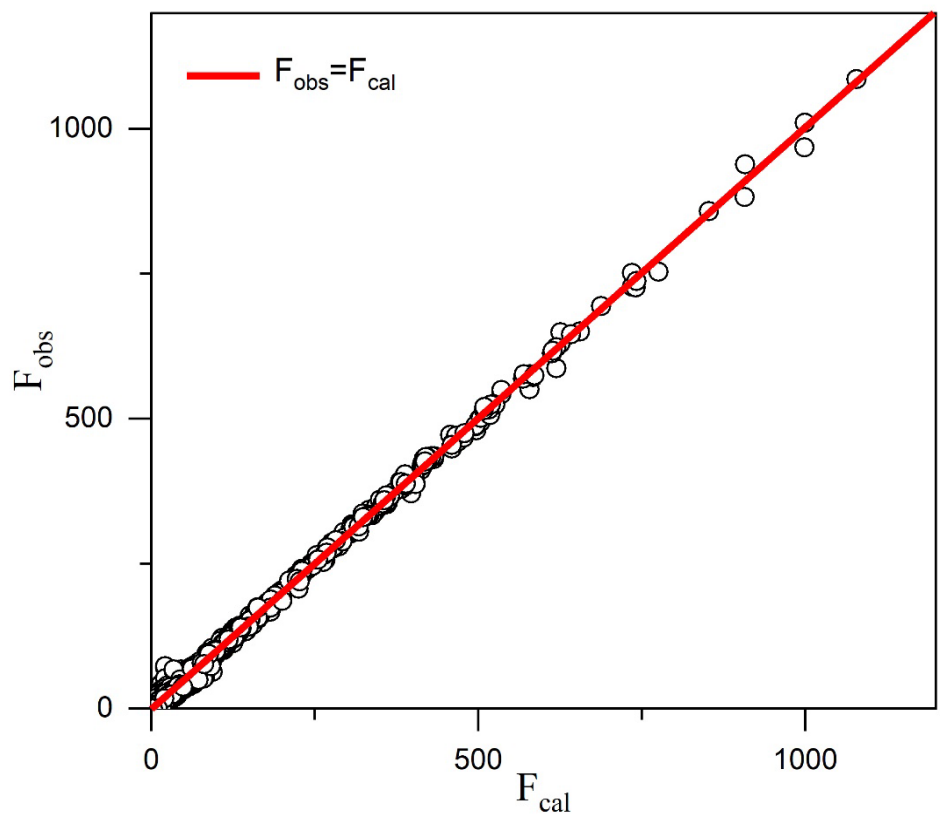

**Figure S3. Calculated structural factor using the *Immm* model as a function of observed structural factor from the single crystal X-ray diffraction data.** The observed structural factor F as a function of calculated F with the red line of $F_{cal}=F_{obs}$. The intensity is proportional to $F^2$.

**Table S1**. Crystal data and structure refinement for La$_2$NiO$_4$·La$_3$Ni$_2$O$_7$.

| Empirical formula | La$_2$NiO$_4$·La$_3$Ni$_2$O$_7$ |
|---|---|
| Formula weight | 1046.68 |
| Temperature | 296(2) K |
| Wavelength | 0.71073 Å |
| Crystal system | orthorhombic |
| Space group | *Immm* |
| Unit cell dimensions | *a*= 5.4296(2) Å |
|  | *b*= 5.4330(2) Å |
|  | *c*= 33.2067(12) Å |
| volume | 979.57(6) Å$^3$ |
| Z | 4 |
| Density(calculated) | 7.097 g/cm$^3$ |
| Absorption coefficient | 26.978 mm$^{-1}$ |
| *F* (000) | 1828.0 |
| Ctystal size | 0.058×0.057×0.040 mm$^3$ |
| Radiation | MoKα (λ = 0.71073 Å) |
| 2Θ range for data collection/° | 4.906 to 72.672 |
| Index range | -9≤*h*≤8, -9≤*k*≤-9, -55≤*l*≤55 |
| Reflections collected | 22488 |
| Independent reflections | 1404 [R$_{int}$=4.11%, R$_{sigma}$=1.70%] |
| Absorption correction | multi-scan |
| Refinement method | Full-matrix least-squares on F$^2$ |
| Data/ restrains/ parameters | 1404 / 0 / 51 |
| Goodness-of-fit on F$^2$ | 1.164 |
| Final R indices [I>2sigma(I)] | R$_1$=6.05%, wR$_2$=10.23% |
| R indices (all data) | R$_1$=7.68%, *w*R$_2$=11.12% |
| Largest diff. peak and hole | 9.96 and -6.29 e. Å$^3$ |

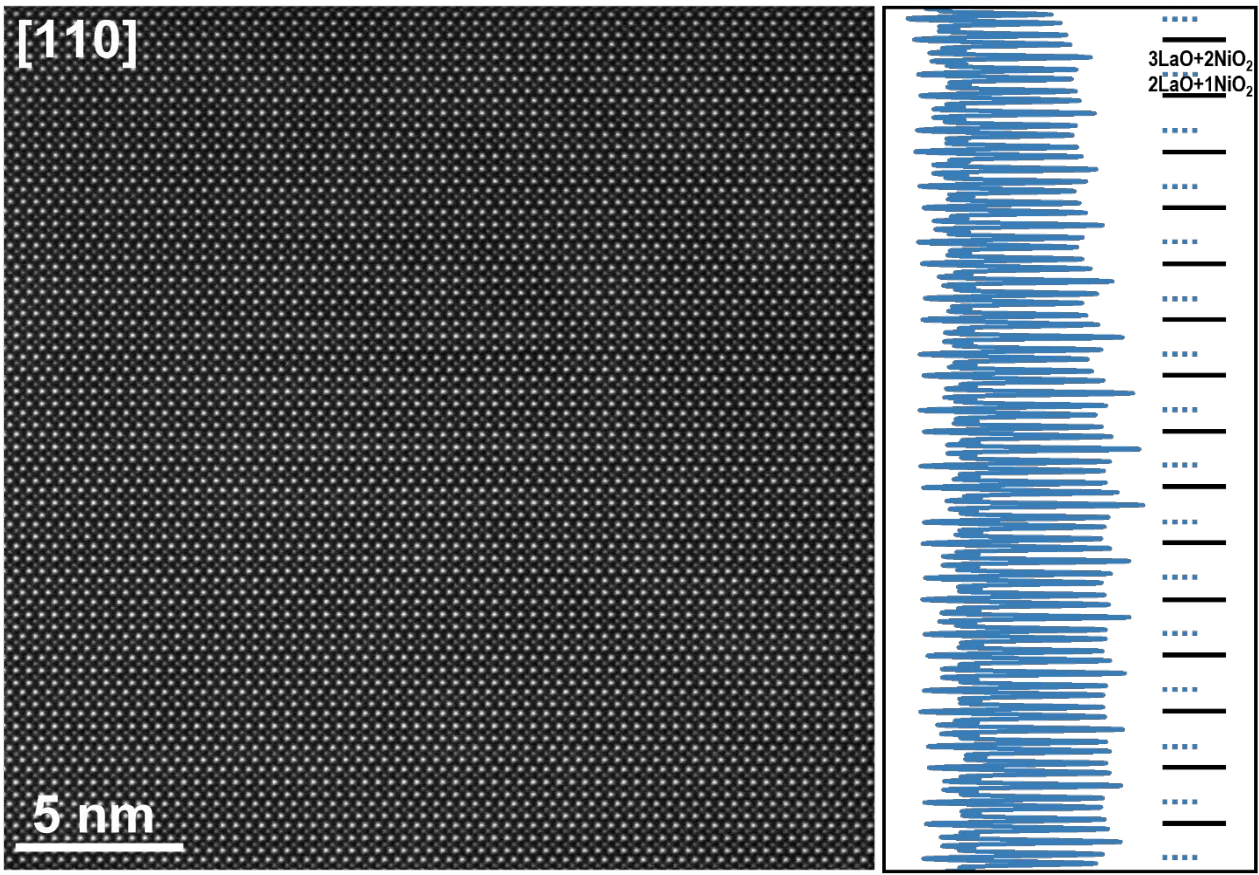

**Figure S4. Real-space imaging of La$_3$Ni$_2$O$_7$·La$_2$NiO$_4$ in the projection of [110].** Lower-magnification HAADF-STEM image in the projection of [110] is shown, the right panel is line intensity profile for rows of all atomic columns. Lower-magnification HAADF-STEM image in the projection of [110] verifies the perfectly ordered stacking of alternating bilayers and single layers.

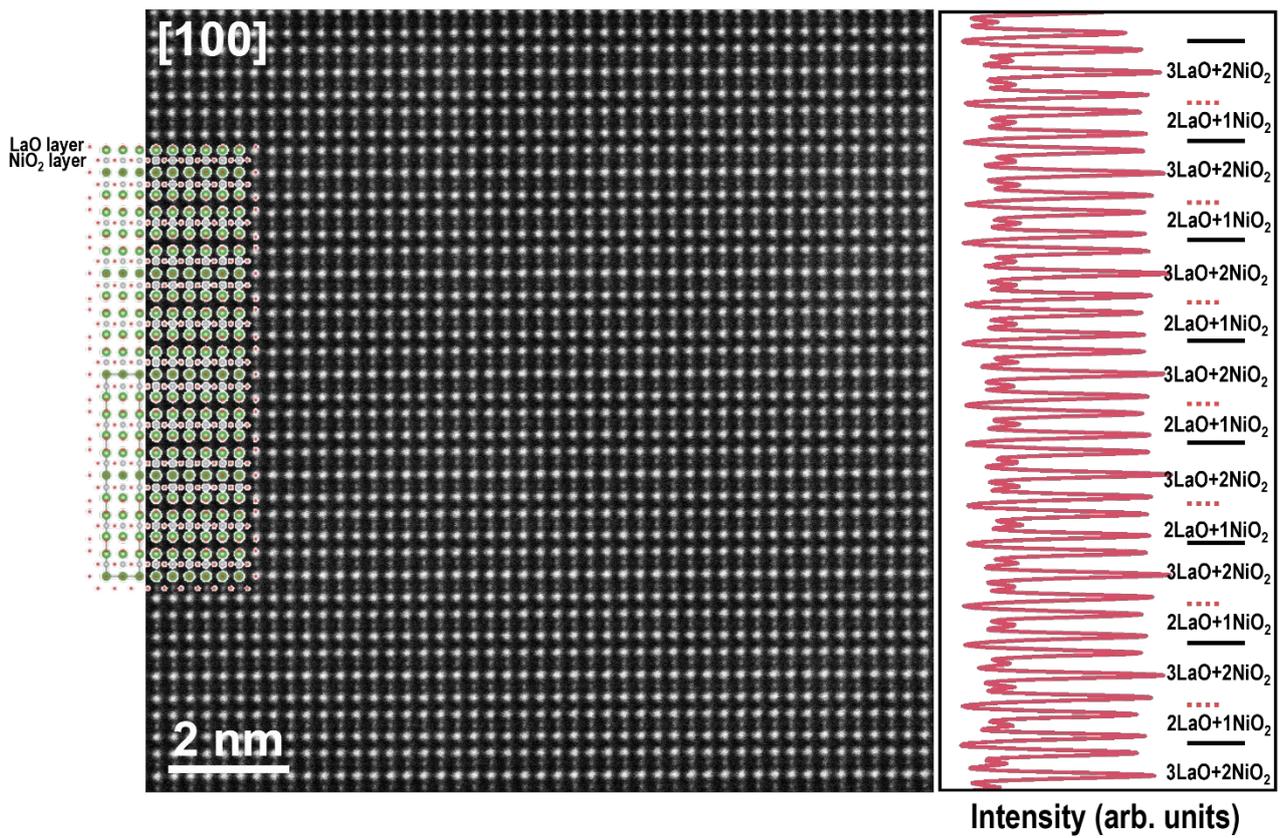

**Figure S5. Real-space imaging of La$_3$Ni$_2$O$_7$·La$_2$NiO$_4$ in the projection of [100].** Stacking of alternating bilayers and single layers is revealed by HAADF-STEM image. The crystal structure model of La$_3$Ni$_2$O$_7$·La$_2$NiO$_4$ is overlaid on the image; the right panel is line intensity profile for rows of all atomic columns.

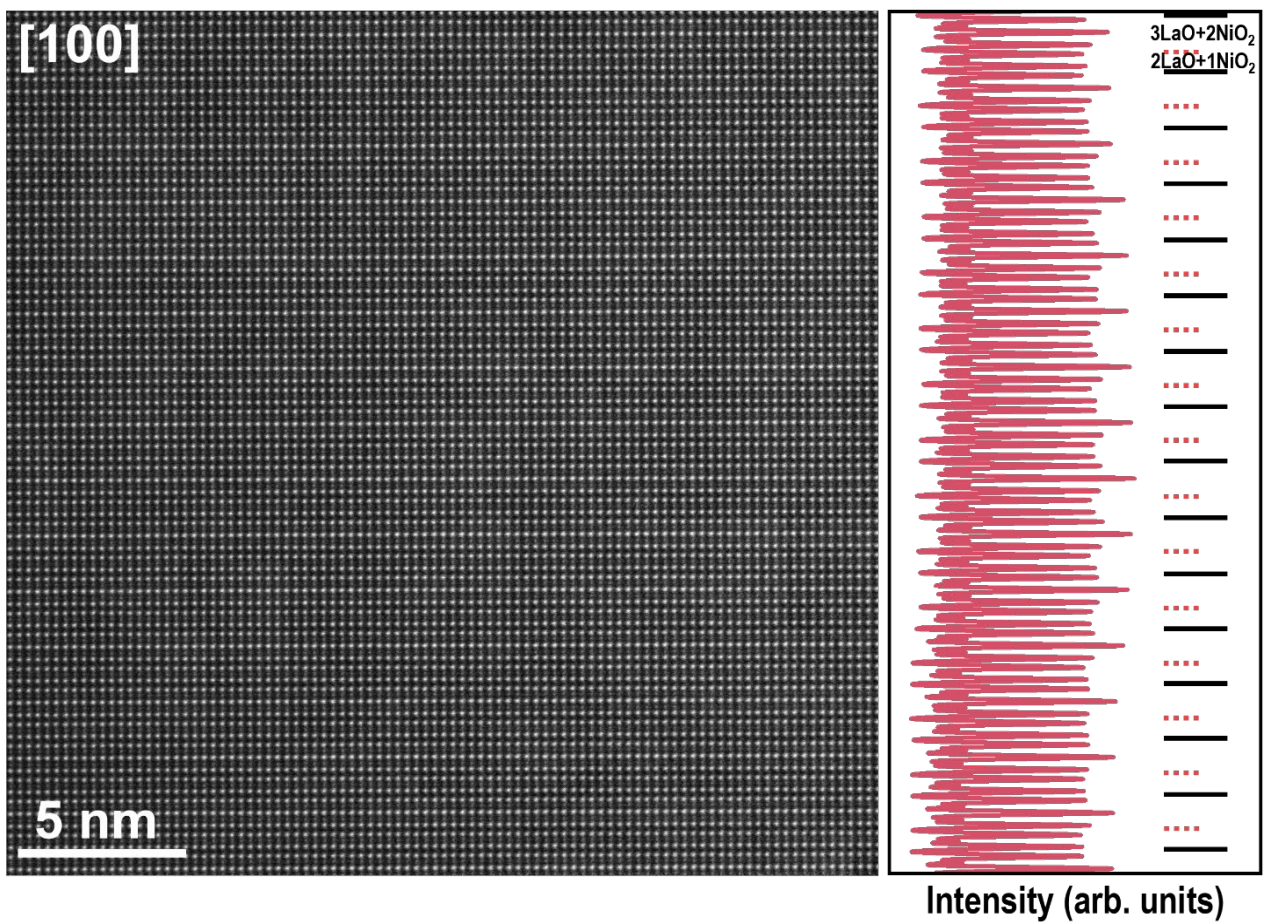

**Figure S6. Real-space imaging of La$_3$Ni$_2$O$_7$·La$_2$NiO$_4$ in the projection of [100] with lower-magnification.** Lower-magnification HAADF-STEM image in the projection of [100] verifies the perfectly ordered stacking of alternating bilayers and single layers. The right panel is line intensity profile for rows of all atomic columns.

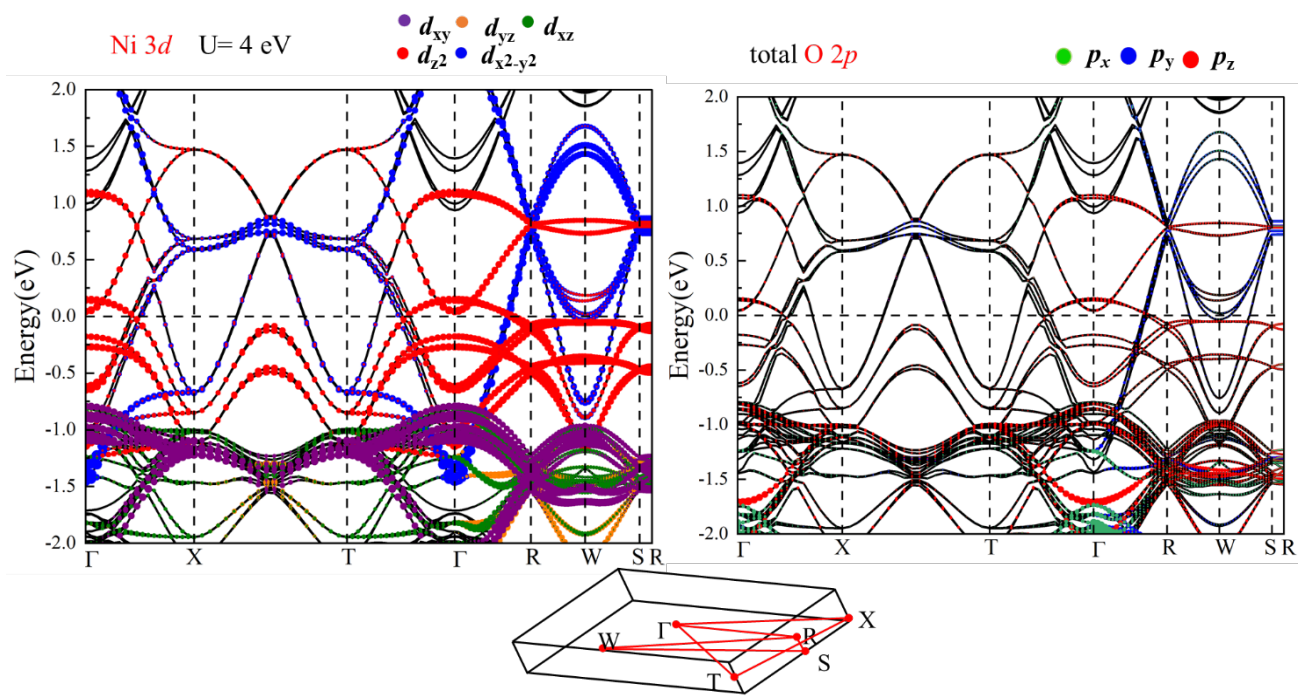

**Figure S7**. DFT+$U$ band structure calculations of non-magnetic La$_2$NiO$_4$·La$_3$Ni$_2$O$_7$. Orbital projected weights from (a) all Ni 3d orbitals and (b) all O 2p orbitals.

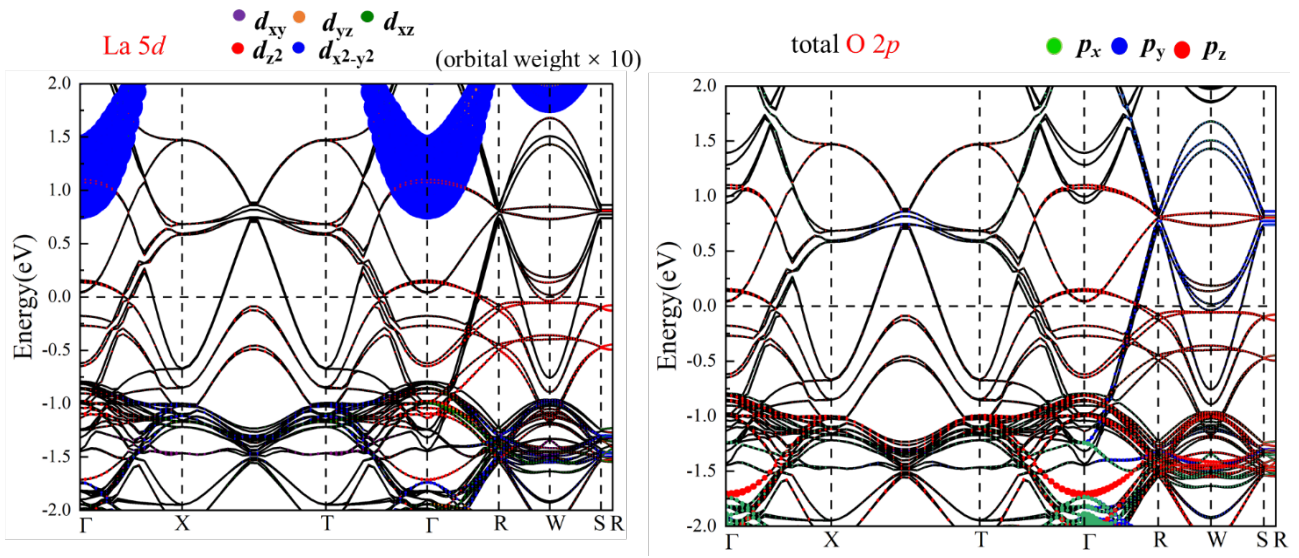

**Figure S8**. DFT+$U$ band structure calculations of non-magnetic $La_2NiO_4 \cdot La_3Ni_2O_7$. Orbital projected weights from La 5d orbitals. Notice that the orbital weight of La 5d orbitals is ten times as large as other orbitals. Compared to O 2p orbitals, it indicates the presence of charge transfer via the LaO chains between the bilayer and mono-layer of nickel oxide planes within the unit cell.